\documentclass[manuscript]{acmart}

\usepackage{multirow}
\usepackage{booktabs}
\usepackage{tabularx}

\acmYear{2021} 
\copyrightyear{2021} 
\setcopyright{licensedothergov}\acmConference[RecSys '21]{Fifteenth ACM Conference on Recommender Systems}{September 27-October 1, 2021}{Amsterdam, Netherlands}
\acmBooktitle{Fifteenth ACM Conference on Recommender Systems (RecSys '21), September 27-October 1, 2021, Amsterdam, Netherlands}
\acmPrice{15.00}
\acmDOI{10.1145/3460231.3474269}
\acmISBN{978-1-4503-8458-2/21/09}


\begin{document}

\title[Follow the guides]{Follow the guides: disentangling human and algorithmic curation in online music consumption}

\author{Quentin Villermet}
\email{quentin.villermet@cmb.hu-berlin.de}
\affiliation{%
    \institution{Computational Social Science Team, Centre Marc Bloch}
    \city{Berlin}
    \country{Germany}}

\author{Jérémie Poiroux}
\email{poiroux@cmb.hu-berlin.de}
\orcid{0000-0002-6948-5574}
\affiliation{%
    \institution{Computational Social Science Team, Centre Marc Bloch}
    \city{Berlin}
    \country{Germany}}
\affiliation{%
    \institution{CNRS}
    \city{Paris}
    \country{France}}

\author{Manuel Moussallam}
\email{mmoussallam@deezer.com}
\orcid{0000-0003-0886-5423}
\affiliation{%
    \institution{Deezer Research}
    \city{Paris}
    \country{France}}

\author{Thomas Louail}
\email{thomas.louail@cnrs.fr}
\orcid{0000-0001-8563-6881}
\affiliation{%
    \institution{CNRS, Géographie-cités}
    \city{Aubervilliers}
    \country{France}
    }

\author{Camille Roth}
\authornote{Corresponding author: roth@cmb.hu-berlin.de}
\email{roth@cmb.hu-berlin.de}
\orcid{0000-0003-3925-7957}
\affiliation{%
    \institution{Computational Social Science Team, Centre Marc Bloch}
    \city{Berlin}
    \country{Germany}}
\affiliation{%
    \institution{CNRS}
    \city{Paris}
    \country{France}}

\newcommand{\tb}[1]{\textcolor{blue}{#1}}
\renewcommand{\tb}[1]{#1}
\newcommand{\fig}[1]{\textbf{\emph{see Figure~#1}}}
\newcommand{\tab}[1]{\textbf{\emph{see Table~#1}}}
\newcommand{\wtf}[1]{\textcolor{red}{\emph{(#1)}}}
\newcommand{\tl}[1]{\textcolor{blue}{Thomas : #1}}
\newcommand{\citex}[1]{\citeauthor{#1} (\citeyear{#1})}
\newcommand{\cnt}{{\small100\%}}

\begin{abstract}
The role of recommendation systems in the diversity of content consumption on
platforms is a much-debated issue. The quantitative state of the art often overlooks the existence of individual attitudes toward guidance, and eventually of different categories of users in this regard. Focusing on the case of music streaming, we analyze the complete listening history of about 9k users over one year and demonstrate that there is no blanket answer to the intertwinement of recommendation use and consumption diversity: it depends on users. First we compute for each user the relative importance of different access modes within their listening history, introducing a trichotomy distinguishing so-called `organic' use from algorithmic and editorial guidance. We thereby identify four categories of users. We then focus on two scales related to content diversity, both in terms of dispersion -- how much users consume the same content repeatedly -- and popularity -- how popular is the content they consume. 
We show that the two types of recommendation offered by music platforms -- algorithmic and editorial -- may drive the consumption of more or less diverse content in opposite directions, depending also strongly on the type of users.
Finally, we compare users' streaming histories with the music programming of a selection of popular French radio stations during the same period. While radio programs are usually more tilted toward repetition than users' listening histories, they often program more songs from less popular artists. On the whole, our results highlight the nontrivial effects of platform-mediated recommendation on consumption, and lead us to speak of `filter niches' rather than `filter bubbles'. They hint at further ramifications for the study and design of recommendation systems.
\end{abstract}

\maketitle{}

\renewcommand{\shortauthors}{Villermet, Poiroux, Moussallam, Louail, and Roth}


\section{Introduction}
The contribution of algorithmic guidance to user preferences and behavior is at the heart of an active debate backed by an \tb{increasing} number of empirical studies. 
This literature typically relies on some notion and measure of diversity, and on some dichotomy between user activity with and without recommendation. 
The latter, diversely denoted as ``organic'', ``autonomous'', ``self-selected'' behavior, generally serves as a reference point of what users would do had they not been \tb{advised or even} diverted by algorithms. \tb{Nevertheless,} many studies focused on user practices have emphasized how user navigation may both passively and actively rely on recommendation devices. The concept of ``navigational queries'' \citep{broder2002taxonomy} first described how search engines may be used to access content that one is already aware of, and where, for instance, autocompletion suggestions merely play a limited role, essentially by reading the user's mind and accelerating their navigation \citep{mitra-2014-on-user}. Beyond this simple old case, \tb{a growing} literature points at the diversity of user practices and expectations toward platform-based recommendation; in the context of music streaming platforms, see  \citep{karakayali2018recommendation,beuscart2019algorithmesen,webster2020taste}. Recommendation most likely does not apply similarly on all users. In other words, its influence should be studied by differentiating user classes who distinctly rely on, and plausibly respond to, recommendation.

This issue should furthermore be framed within a broader context where guidance may stem both from algorithmic devices and from human recommendations. The latter is arguably not new: friends and relatives, mainstream media and salesclerks have traditionally played a prime role in \tb{suggesting and} shaping consumption preferences. In the music realm, the composition of radio playlists has typically long been decided by humans. 
The \tb{flourishing development} of human-mediated, or ``editorial'', guidance \citep{bonini-2019-first,bandy2020auditing}
may however sometimes appear as a blind spot \tb{in the appraisal of} platform-based recommendation.
Building upon this recent trend, we extend the usual algorithmic/organic dichotomy by considering editorial guidance as a specific, distinct mode of recommendation. 
Contemporary music streaming platforms epitomize this trichotomy for they feature the three types of access to content: \emph{algorithmic} (songs suggested by a recommendation engine, for instance in autoplay mode), \emph{editorial} (songs curated by the platform staff, for instance as ``featured playlists'') and \emph{organic} (self-selected songs, either from personal playlists and favorite albums, or directly sought through the search bar). \tb{Note that in all three cases, it is obviously the user who chooses the mode of access: what matters in this trichotomy is that song titles are  proposed by an algorithm, the editorial staff, or the user themselves.}

The case of music streaming further 
makes it empirically feasible to connect \tb{the role of platform-based recommendation with traditional} content curation. \tb{Specifically, we are able to contrast listening practices of users who rely more or less on some types of platform recommendation affordances with} the mass (and somewhat traditional) guidance proposed by music radios, as an \tb{offline or ``off-platform''} reference point. Our study is thus based on two datasets: first, anonymized yet comprehensive listening histories for several thousands of users active on a major music streaming platform over a certain period of time. Second, playlists for a selection of mainstream radios in France over the same period. 

\smallskip
On the whole, we aim to concretely examine (i) whether there exists distinct behavioral patterns with respect to the three modes of access to content and (ii) whether some type of guidance tends to pull listeners towards exploring certain types of content. We characterize the exploratory nature of listening histories with two distinct notions, dispersion and \tb{artist popularity}, which correspond to two dimensions of diversity --- i.e., dispersion at the user level  (skewed popularity of content within a given user's listening portfolio: redundancy or not) and dispersion at the macro level (skewed popularity of \tb{artists} over all users: mainstream or not).

\smallskip
The main contributions are as follows:

\begin{itemize}
\item we hypothesize the existence of distinct classes of user attitudes towards recommendation before examining its possible influence on content consumption patterns;
\item we further differentiate between two types of platform-side recommendation, stemming either fully from algorithmic suggestions or primarily from human curation;
\item we establish a bridge between online platform recommendation and one of the traditional types of music guidance from the non-platform world, exemplified by radio music playlists.
\end{itemize}

\section{Related work}

\paragraph{Diversity and recommendation systems.}
The role of recommendation devices in fostering diversity is assuredly at the heart of a fast-growing literature \tb{focused on a myriad of platform types, where music streaming is only one application case among many}. Contrarily to popular assumptions about so-called ``filter bubbles'', the emerging empirical picture suggests that recommendation algorithms generally seem to increase diversity and serendipity
\citep{baks-expo,haim-2018-burst,aiel-evol,moller-2018-do-not-blame,puschmann2019beyond,roth-2019-algorithmic}, even though recent results on specific platforms such as Spotify or YouTube tend to suggest otherwise \citep{anderson2020algorithmic,roth2020tubes}, while explicit personalization or ``self-selection'' also appear to induce algorithmic reinforcement and confinement\tb{, for instance regarding news consumption} \citep{zuiderveen-borgesius-2016-should,DYLKO2017181}. 
Most of this literature works at the aggregate level without distinguishing populations of users who may differently use or respond to algorithmic guidance. \tb{Several} studies nonetheless expressly differentiate users who are eager for recommendation \citep{nguyen2014exploring}, diversity \citep{munson-2010-presenting}\tb{, or exploration \citep{Gathright-Understnading-2018,Hosey-JustGiveMe-2019}}. 
This hints at the existence of different user behaviors and expectations towards recommendation \citep{karakayali2018recommendation} prior to it influencing users.

\paragraph{Use of music streaming platforms.}
Music streaming platforms feature an array of recommendation uses. They may be employed as ``intimate experts'' \citep{karakayali2018recommendation} who are able to counsel users in the most personalized manner and contribute to the self-formation of taste. They also provide editorial guidance which induces a hybrid type of curation, distinct from pure algorithmic recommendation \citep{bonini-2019-first}. 
The co-existence of algorithmic and editorial guidance is not unique to music streaming: their differential effect has already been studied on news platforms, where editorial recommendations appear to outperform algorithms in terms of diversity or \tb{in terms of} concentration \citep{bandy2020auditing}. 
This issue connects with a relatively older literature appraising the role of online \tb{retail} platforms in the consumption of content of varying popularity, especially at the top and bottom of the so-called ``long tail'' \citep{elberse-2008-should,goel-2010-anatomy}. Curation on music platforms \tb{also} opens specific opportunities with respect to the access to niche content \citep{barna2017perfect}. In this regard, quantitative studies show that platform use in general is associated with listening to less mainstream and less redundant content \citep{datta2018changing}, even though diversity seems to plateau for the most active users \citep{poulain2020investigating}. However, the proper contribution of recommendation in these aggregate studies remains difficult to assess. Two recent studies differentiate organic from guided navigation and appear to yield conflicting results, whereby the former \citep{anderson2020algorithmic} or contrarily the latter \citep{beuscart2019algorithmesen} is shown to be correlated with increased \tb{consumption} diversity. Notwithstanding, here again, little is known on the existence of underlying types of users (or even classes of users in the sociological sense, as in \cite{webster2020taste}) who make use of the platform affordances in different ways --- and, in turn, how the same affordances distinctly affect these user populations in their exploration of different regions of the musical space.

\paragraph{Music recommendation goals.}
The intertwinement between expectations, uses and influences is not foreign to the more technical issue of the design of recommendation. 
The introduction of recommending devices is generally posterior to the creation of music streaming platforms, which were \tb{often} initially operating as pure digital libraries \tb{(a notable exception being Pandora)}. While recommendation may first have been construed as an \tb{``attentional trap'' \hbox{i.e.,} to help retaining the user longer on the platform} \citep{seaver2019captivating}, the question of what it aims to optimize and how it intends to improve user experience provides further context to our contribution. \tb{The work of} \citep{bonnin2014automated} offers a comprehensive overview of the current techniques for automated playlist creation, and their evaluation. \tb{It emphasizes} in particular the importance of taking into account song popularity \tb{(\hbox{e.g.,} to tackle the cold-start problem \citep{bauer2019global})}, while radio playlists are said to typically ``contain popular tracks in a relatively homogeneous style''.  Models that generate and rely on categories of user tastes have also been proposed \tb{\citep{zheleva2010statistical,vigliensoni2016automatic}} and seem to go much further than the above-mentioned literature in framing algorithmic recommendation in terms of implicit user classes.
Regardless, the conception of music recommendation systems already makes significant use of so-called ``beyond accuracy'' measures \citep{mcnee2006being} by attempting at integrating diversity-related metrics \cite[variously denoted as coverage, serendipity, novelty, see][]{zhang2012auralist,kaminskas2016diversity,schedl2018current}. How these optimization principles may be translated, integrated and combined into platform-wide practices largely remains an open and active question.

\section{Dataset}
We obtain song listening histories from Deezer, a French private music streaming platform which started operations in 2007. Deezer currently has about 14m monthly active users worldwide, about half of whom are paid subscribers as of January 2019. Our data set describes twelve months of anonymized activity over 2019 for 8,639 randomly chosen paying subscribers based in France who were registered to the service before December 31st, 2018 and active in January 2019. \tb{We limit ourselves to users with a paid subscription, even though a different listening behavior of users of the free service might be assumed from previous work \citep{wlomert2016demand}.}
We discard events lasting less than 30s, assuming these are so-called ``skips''. The dataset ultimately totals about 51m timestamped streaming events (plays) describing which user listened to which song by which artist for how long \hbox{i.e.,} 5.9k plays per user on average (or around 16 per day).

The data also indicates which product feature users accessed songs through. Note that the platform may be indifferently used from a dedicated desktop app, a mobile app, or directly from a browser: the interface generally provides the same functions irrespective of the chosen device.  On Deezer, users may directly look for music titles, albums or artists using a search bar. 
They can tag songs, artists and albums as favorites and build their own playlists. We denote these modes of access as ``\emph{organic}'' for they entirely rely on user choices, whereby users do look for a specific item in their or the platform's library. Users can also navigate to a home page where they find tailored recommendations which are either assembled and labeled by human editors at Deezer (such as recommended playlists variously called ``10s electronic'', ``Rock \& Chill'', etc.) or algorithmically curated (such as the so-called ``flow'', which is a personalized automatic mix). We denote the former as ``\emph{editorial}'', since the content is curated and recommended by human editors, and the latter as ``\emph{algorithmic}'', for it entirely relies on an interaction between Deezer algorithms, platform-wide data and users' listening histories. In some cases, editorial playlists are algorithmically selected to be presented to users. We classify these as ``\emph{editorial}'' since content selection remains primarily the choice of human editors. There is admittedly some porosity in the catalogs related to each of these three modes: for one, a user may include an algorithmically- or editiorially-discovered song into one of their playlists; similarly, algorithmic recommendation may feature songs a user already listened to organically. 
\tb{Transfers between various access modes have been analyzed in \citep{shakespeare2021}. We focus on the properties of song listening portfolios associated to such and such type of user and affordance.  Combining this analysis with transfers would make it possible to characterize adoption and influence dynamics in the mid- to long-term but remains beyond the scope of the present paper and is left for future research.} 
Whenever a song is listened to through some access mode, we consider that it is because the user decided to rely on that access mode, and thus describes listening patterns associated to that mode, irrespective of their more distant origin.

\smallskip
We also collect comprehensive playlist histories for a selection of 39 mainstream radios in France over the same period. This includes the top 15 French national musical stations in terms of measured audience during 2019, along with a relatively arbitrary selection of more specialized stations and webradios. For each radio, broadcast logs were collected and songs matched against Deezer's whole catalog. 

\section{User practices}

\subsection{Modes of access and user behavior classes}
The \tb{quantitative} literature \tb{does not often pay} attention to the possible existence of distinct classes of users when it comes to their behaviors and listening habits on the platform, classes for which the function and effect of recommendation may differ significantly \tb{\citep{Gathright-Understnading-2018}}. \tb{Many} studies report aggregate effects averaged over binary categories of users, for instance depending on a heavy \hbox{vs.} limited use of recommendation \cite{nguyen2014exploring,datta2018changing} or categorical variables such as gender \cite{shakespeare-2020-exploring} or age \cite{anderson2020algorithmic}. We assume (and confirm) that a pre-grouping of users depending on broad use classes may reveal distinct sensitivities to recommendation. In other words, we contend that users who, say, rely more on editorialized playlists may respond differently to algorithmic recommendation than users who are mainly organic.

\begin{figure*}[t]
  \centering
  \includegraphics[width=.83\linewidth]{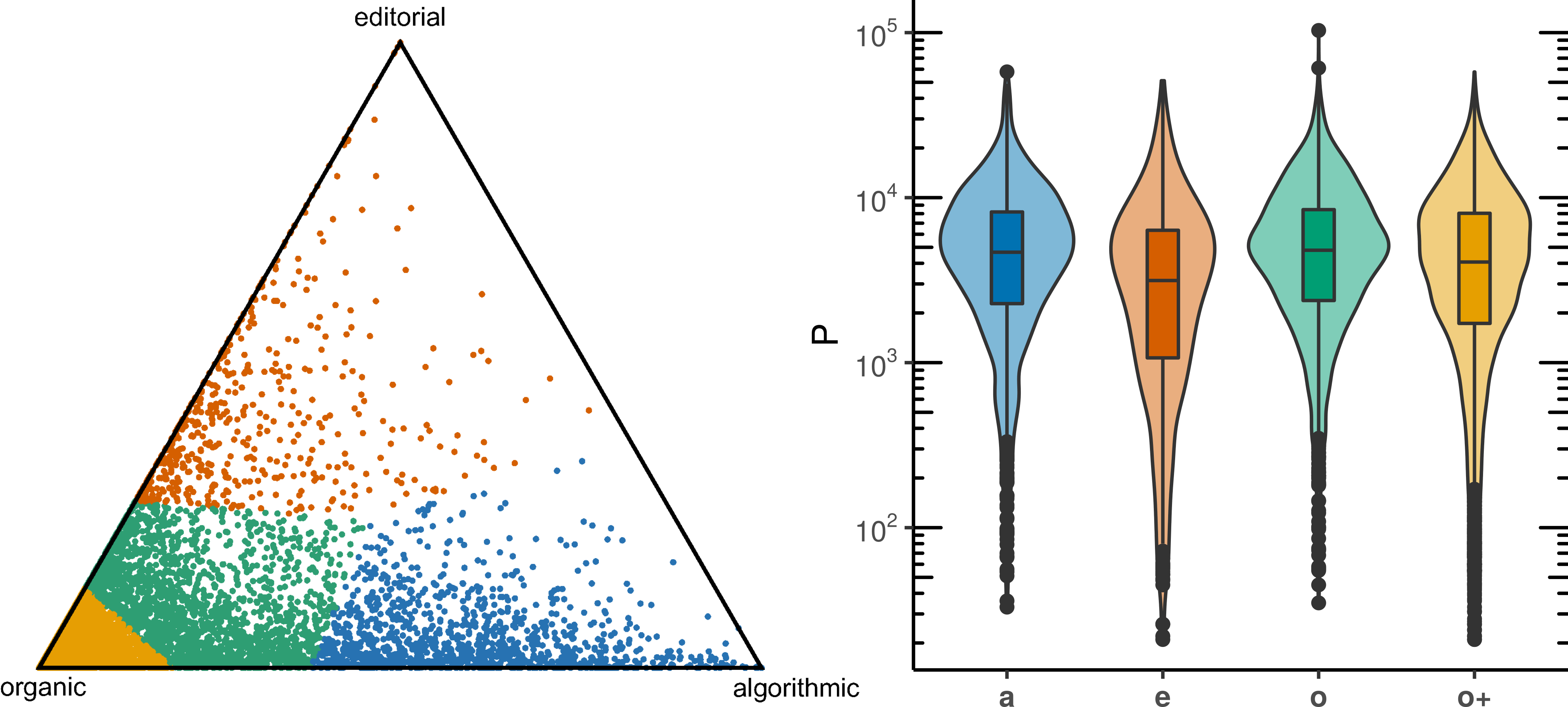}
  \caption{{\bf Left:} Use profiles and classes, where each dot on the ternary plot corresponds to a user of barycentric coordinates $(p_a,p_e,p_o)$, and each color refers to one of the four classes {\bf a} (blue), {\bf e} (red), {\bf o} (green), {\bf o+} (yellow). {\bf Right:} Box and violin plots of activity (number of plays $P$, log-scale) for each user class.\label{fig:binu_distp}}
  \label{fig:kmeans_ubins}
   \Description[Shows the partition into four user classes in a ternary space]{Left: Users are shown as colored dots in a ternary space and segmented in 4 classes, each color corresponds to a class. Right: distribution of activity for the 4 user classes look similar}
\end{figure*}

Let us denote $P$ the number of plays in a user's listening history, of which proportions $p_a$, $p_e$ and $p_o=1-(p_a+p_e)$ have respectively been accessed algorithmically, editorially and organically. Their access mode profiles may thus be described by a triplet $(p_a,p_e,p_o)$ defining barycentric coordinates in a ternary space, as shown in \tb{Figure}~\ref{fig:kmeans_ubins}-left. Even though a significant portion of users visibly rely to a large extent on the organic mode, we nonetheless observe a great deal of heterogeneity which hints at distinct use behaviors. We further define user groups by performing a simple Voronoi partitioning of profiles using a $k$-means algorithm. Choosing $k=4$ explains around 80\% of the variance with limited improvement for $k>4$. Other clustering methods have been tried but generally yield a single cluster, most likely because data density increases roughly monotonously in the direction of the ``organic'' vertex, thereby preventing the formation of marked boundaries. As such, this partitioning only aims at defining bins as areas of the ternary space, rather than well-separated clusters per se. \tb{As a result, our four clusters should be understood as a simple approach to create an independent variable of use profiles among and between which we observe marked behavioral heterogeneity.} It is similar, albeit not equivalent, to defining, say, \tb{quartiles or} deciles of income, which notoriously obey a hybrid log-normal/Pareto distribution with no clear-cut boundaries. 

In the following we thus consider four user bins that we denote as ``{\bf a}'' (rather ``algorithmic'', 989 users), ``{\bf e}'' (rather ``editorial'', 655 users), ``{\bf o}'' (rather ``organic'', 1614 users) and ``{\bf o+}'' (``very organic'', 5381). To avoid any ambiguity with the various other practical bins that we define and use later on, we will now denote these user bins as ``\emph{user classes}''.
On the whole, organic classes ({\bf o} and {\bf o+}) comprise 7062 users \hbox{i.e.,} roughly 80\% of the dataset, for whom at least half of all plays have been accessed autonomously by users. Table~\ref{tab:centroids} gives the shares of plays in each of the three modes for the centroids of these four classes. The number of plays $P$ is also a proxy of the user's activity on the platform. Its distribution spans several orders of magnitude, in a manner similar for all classes, as can be seen in \tb{Figure}~\ref{fig:binu_distp}-right --- in other words, there are weakly and strongly active users in all classes. We note that average activity is nonetheless slightly smaller for ``editorial'' users.

\begin{table}[]
    \centering\small
    \begin{tabular}{lrlp{.0cm}ccc}\toprule
        \em user& \multicolumn{2}{c}{\em population} && \multicolumn{3}{c}{\em access mode share}\\
        \em class&\#users&(\% dataset)&&organic& algorithmic & editorial \\
        \midrule
        \bf a& 989 &(11\%) &&36\% & 58\% & \:\:6\%\\
        \bf e& 655 &(\:\:8\%)&&55\% & \:\:7\% & 38\%\\
        \bf o& 1614 &(19\%)&&70\% & 23\% & \:\:7\%\\
        \bf o+& 5381& (62\%)&&94\% & \:\:2\%& \:\:4\%\\\bottomrule
    \end{tabular}
    \caption{\tb{Population of each user class and proportion of plays by access mode for users chosen as class centroids} (rows sum to 100\%).}
    \label{tab:centroids}
    \Description[Population sizes of each user class (fully described in text) and respective shares of access mode, which vary significantly]{Organic access is the primary access mode for all user classes except the algorithmic ones ``a'', editorial access is marginal for all but users labeled as editorial ``e''. Difference between organic ``o'' and organic+ ``o+'' is the level of predominance of the organic access mode.}
\end{table}

\subsection{Two dimensions of diversity}

We rely on these four classes to appraise the role of each access mode on music consumption, in terms of \emph{where} each mode brings users to, and \emph{how much}. We characterize the portfolio of songs the users listen to by focusing on two fundamental diversity measures:
\begin{enumerate}
\item {dispersion}, denoting the lack of redundancy in the listening history;
\item \tb{{artist popularity}}, denoting the tilt toward \tb{songs by more popular} artists.
\end{enumerate}
We explain below (\ref{sec:mainstreamness}) why we chose not to use genre meta-data to appraise diversity.

\subsubsection{Dispersion and activity.}
\begin{figure}[t]
\centering
\includegraphics[width=.415\columnwidth]{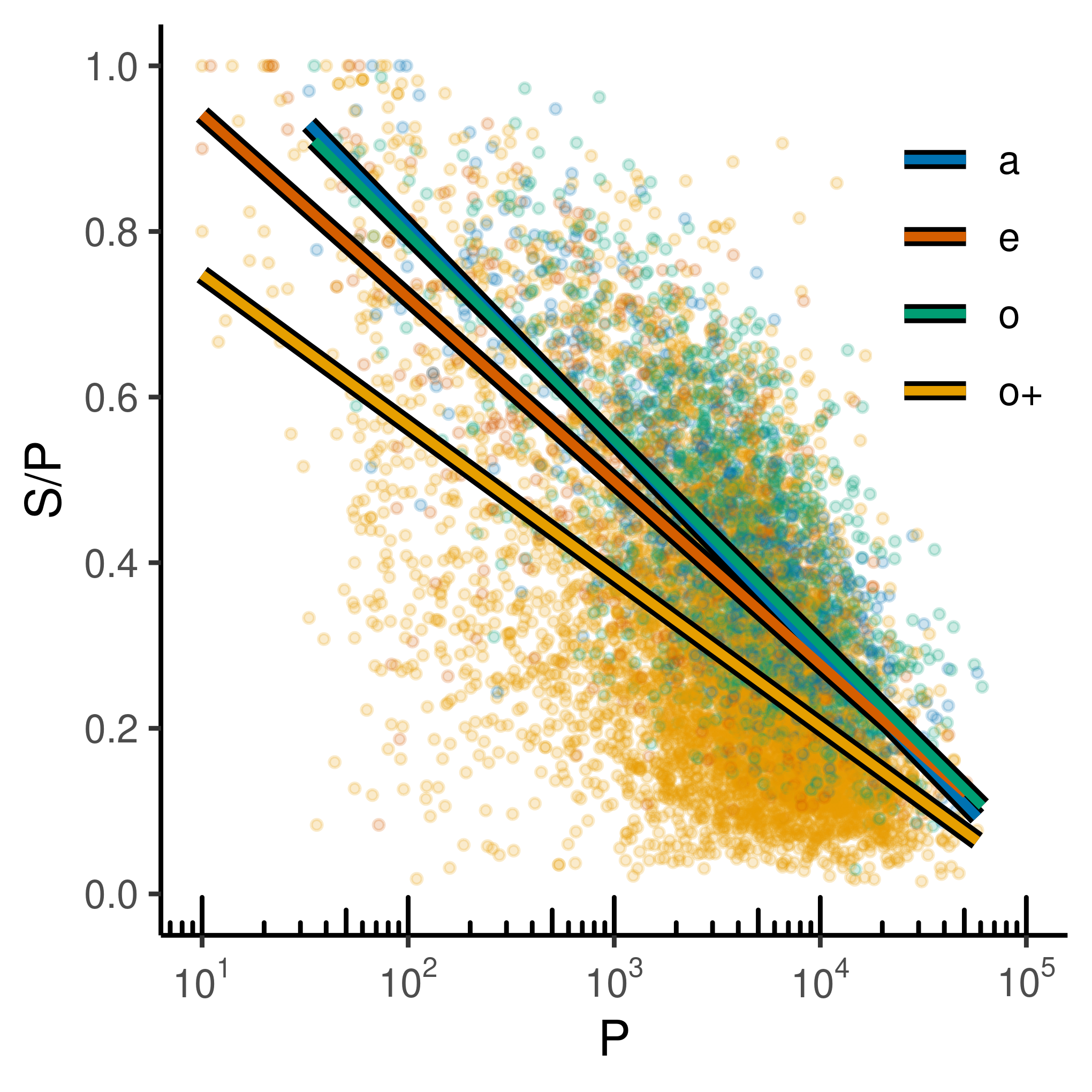}
\includegraphics[width=.415\columnwidth]{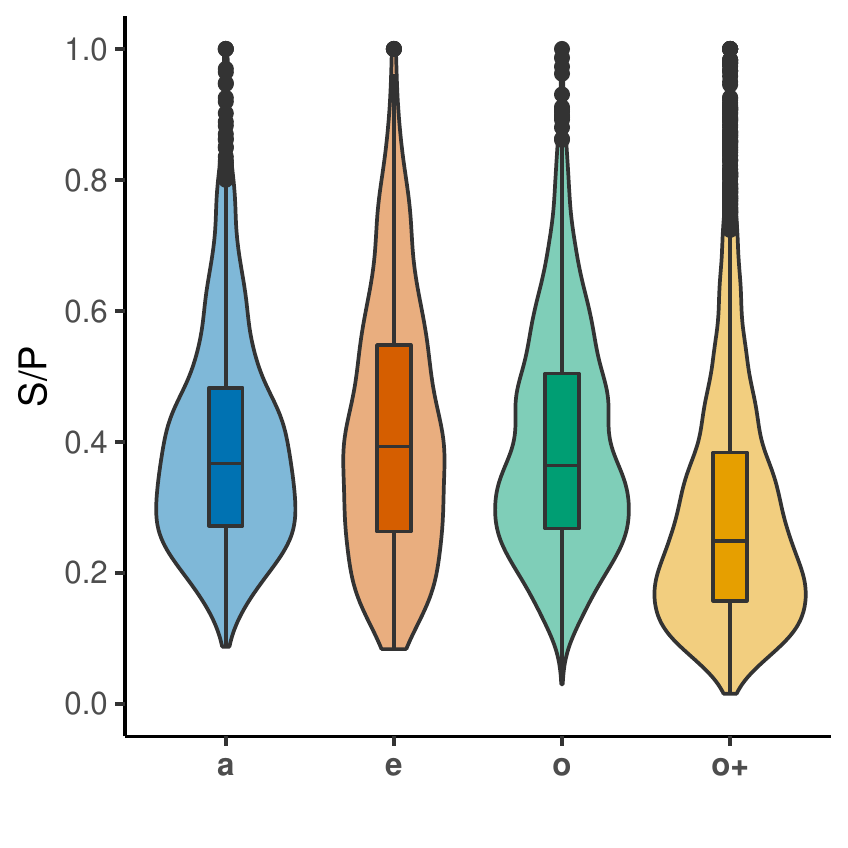}
\caption{Dispersion profiles for each user class. {\em Left:} scatterplot (each dot is a user) and binned averages (solid lines) as a function of activity ($P$, log-scaled). {\em Right:} boxplots per user class.}
\label{fig:evoSP}
\label{fig:sp_ubins}
\label{fig:sp_logp_binu}
\Description[Scatterplot and distributions of dispersion profiles]{On the left, users from the four classes are scattered in the diversion versus activity plane but organic users have on average lower dispersion for a given activity level. This can also be seen on the distributions per class on the right.}
\end{figure}

{Dispersion} may simply be computed as $S/P$, where $S$ is the number of unique songs among the $P$ plays. It thus equals $1$ when each song is played exactly once, and goes to 0 as plays of the same songs are repeated. \tb{In other words,} low dispersion implies high redundancy \tb{of a listening history}. $S/P$ is also called ``exploratory ratio'' in \citep{louail2017headphones} and \tb{further} alludes to the trade-off between exploration and exploitation. \tb{Whichever the preferred interpretation, this} ratio relates to a \emph{functional} understanding of the diversity of a user's listening behavior.
While the data display large fluctuations, we fit for each user class a linear model to highlight the general tendency. This reveals an inverse relationship between dispersion and activity, for all user types (see \tb{Figure}~\ref{fig:evoSP}-left): a user's catalog tends to saturate and its \tb{dispersion decreases with listening time (put differently, redundancy \hbox{i.e.,} exploitation of the catalog increases}). For a given activity level, dispersion is nevertheless lower for very organic users, while it is comparable for moderately organic and for recommendation-intensive users; \hbox{i.e.,} as soon as some form of recommendation is involved. This indicates a general inclination for redundancy with users who rely most on organic access modes. Focusing on recommendation-intensive users, dispersion is slightly lower for editorial users. However, their overall lower activity, which in turn corresponds to higher dispersion, leads to levels of dispersion comparable with those of algorithmic users, as shown on \tb{Figure}~\ref{fig:evoSP}-right. 

\begin{figure*}[t]
\centering
\includegraphics[width=.92\linewidth]{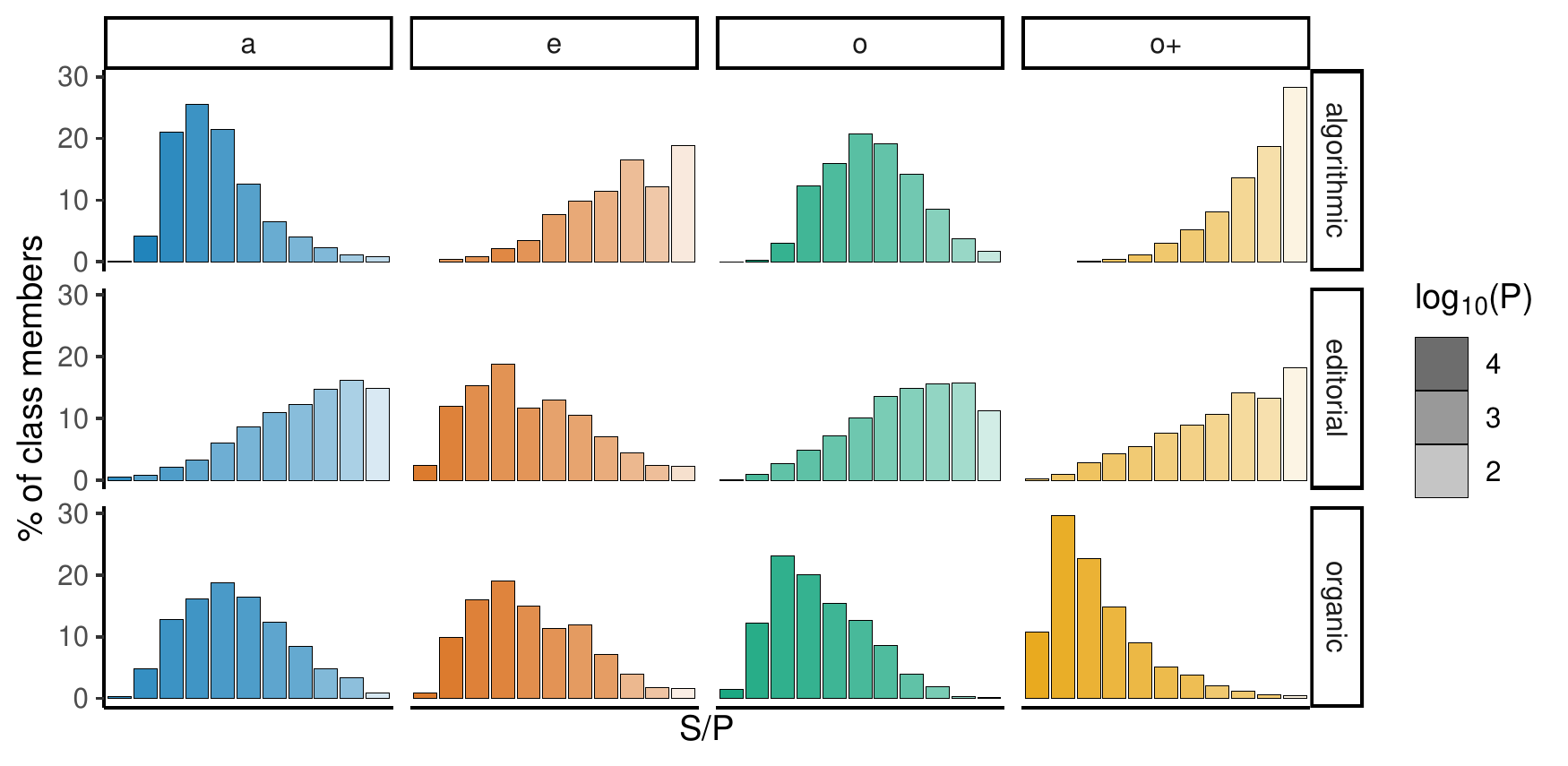}
\caption{Breakdown of dispersion values for each user class and for each access mode. Histograms are further binned by deciles of increasing $S/P$ values (from 0 to 1 from left to right) and indicate how many users of each class ({\bf a}, {\bf e}, {\bf o} and {\bf o+}) exhibit which dispersion value for a certain access mode (algorithmic, editorial or organic). Average activity values for users of each decile bar are further indicated by a grayscale, where darkest shades correspond to highest $P$ values.}
\label{fig:pdist}
\Description{Figure 3: fully described in the text}
\end{figure*}

We refine this picture by breaking down dispersion values per access mode. Figure~\ref{fig:pdist} reveals that, for all user classes, the major access mode generally exhibits lower dispersion values --- all the more so when activity is higher for a given access mode, as shown by the fact that the darkest bars (i.e., highest activity) are generally to the left of these histograms (i.e., lowest dispersion). Put simply, for algorithmic users ({\bf a}), \tb{dispersion is lower for algorithmic recommendations and higher} for editorial ones. For editorial users ({\bf e}), dispersion is lower for editorial recommendations\tb{, and so on:} the same applies to organic ({\bf o}) and even more so for very organic users ({\bf o+}) who exhibit \tb{high dispersion in both algorithmic and} editorial access. 

This suggests a phenomenon of specialization in each class of users, who tend to prioritize exploitation in their preferred mode of access, and favor exploration in the others, in terms of \tb{more dispersed} plays. We shall here keep in mind that very organic users exhibit the \tb{lowest dispersion} --- using the platform heavily as a \emph{digital library} connotes more exploitation of the catalog than using it as a \emph{digital librarian}. In other words, even a moderate use of some form of recommendation is generally associated with a higher level of exploration.  

\subsubsection{\tb{Artist popularity} and access modes.}\label{sec:mainstreamness}

\tb{Artist popularity}, by contrast, describes music consumption \tb{in terms of content created by popular or less popular artists}. We denote it as $\nu$. It relates to a \emph{semantic} understanding of diversity. 
The literature traditionally proposes two main approaches to describe such diversity. On the one hand, some studies appraise diversity in spatial terms by hypothesizing an underlying ``cultural'' or music similarity space, and by embedding songs \citep{west2006model,airoldi2016follow,anderson2020algorithmic}, individuals \citep{lambiotte2005uncovering,savage2011unravelling} or both \cite{hsieh2017collaborative} into a graph or a low-dimensional space where proximity roughly corresponds to semantic similarity (e.g., similar musical genres or tastes). Such spaces are also used by platform algorithms to recommend songs not too far from one another (e.g., avoiding recommending manouche jazz after a heavy metal song). 

On the other hand, some studies characterize consumption diversity through content popularity, most notably through the share of top artists \citep{datta2018changing,beuscart2019algorithmesen}. 
We follow here this latter approach for three main reasons.

First, designing an embedding space requires choosing between many possible similarity metrics and construction approaches, each based on strong hypotheses on the semantic encoding. For instance, distances based on the co-occurrences of content tags (e.g.  taxonomies \cite{hennequin2018audio} or folksonomies \cite{park2015understanding}), will typically suffer from definition ambiguities and socio-cultural inconsistencies \cite{beer2013genre,epure2020modeling} and more prosaically from annotation noise and biases. Distances based on acoustic similarities \cite{van2013deep} or usage \cite{hsieh2017collaborative} are typically tailored for neighborhood consistency, but will not generally bear any meaning for, say, the distance between metal and jazz songs, compared with the distance between jazz and classical pieces.
In addition, many users, particularly omnivores \citep{peterson1996changing,van2001social} may show interest for more than one genre. Navigation profiles may thus exhibit several distinct centroids in possibly distant regions. In turn, this potential ``musical polycentrism'' might blur the meaning of notions solely based on the geometric extent of listening profiles in multi-dimensional spaces. 
\tb{Conversely, \textit{artist popularity}} remains a mono-dimensional notion, whose mean and deviation are simple yet likely robust indicators of the position and span of a user's musical content consumption with respect to the whole field. 

Second, this will ease the comparison with radio playlists and thus offline editorial recommendation. The radios of our dataset indeed address a quite diverse collection of music genres, some being very generalist (\hbox{e.g.,} `FIP'), some much more specialized (\hbox{e.g.,} `TSF Jazz'). To discuss specialization and eclectism across genres and radios, we believe that \tb{artist popularity} may act as a better sort of \emph{lingua franca} than, say, extents in a vector space loosely encoding cultural similarities. 

Third, \tb{artist popularity} directly connects with an older debate on whether the almost infinite catalogs of online platforms foster consumption of more mainstream or more niche content, or both, and by which types of users \citep{elberse-2008-should}. 

We first define \tb{artist popularity} by computing how many times their songs have been played in the whole dataset as a proxy of \tb{their} popularity on the platform. \tb{This is referred as ``artist playcounts'' in \citep{bauer2019global}, also connected with the mainstreamness of artists.}
We then distinguish four bins of popularity such that each bin gathers a similar numbers of plays i.e., artists which, taken together, represent the same total amount of plays $(\frac{1}{4}\sum_\text{users} P)$. This ensures that a play chosen in a uniformly random way from the listening history has the same likelihood of belonging to any of the four bins. As a result, the first bin $\nu_1$ gathers the top 73 artists, the fourth bin $\nu_4$ the bottom 166,869 artists. 

Table~\ref{tab:binmu_types} gathers the proportion of access modes for each \tb{popularity} bin. In general, songs are principally listened to organically, irrespective of \tb{artist} popularity --- around 80\% of all plays on average, which is assuredly expected given the prevalence of organic access modes (figure~\ref{fig:kmeans_ubins}). 
We however see that organic access is non-monotonous, in that the highest shares are found in both the most and least \tb{popular} artist bins \citep[consistent with][]{goel-2010-anatomy}.
A different picture emerges for guided access modes. Algorithmic access puts the highest share on intermediate bins, and by far the lowest share on the \tb{highest popularity} bin. Editorial access, by contrast, is monotonously more frequent for \tb{higher popularity} bins. 
In other words, \tb{content by popular artists} seems to be more often proposed by editorial rather than algorithmic picks. 
To further exhibit this effect, we compute the average \tb{popularity} bin for each access mode and find 2.68 for algorithmic plays, 2.49 for organic ones, and 2.23 for editorial ones (the average \tb{popularity} bin for all plays is 2.50, by construction). 

\begin{table}[t]
    \centering
    \begin{tabular}{r>{\small}r|ccc|>{\scriptsize}c}\toprule
    		\multirow{2}{*}{bin}&\multicolumn{1}{c|}{\multirow{2}{*}{\# artists}}&\multicolumn{3}{c}{\em access mode}\\
		&\multicolumn{1}{c|}{}&\small algorithmic&\small editorial&\multicolumn{1}{c}{\small organic}\\\midrule
        $\nu_1$	&	73	&9\%		&\bf8\%	&\bf83\%	&100\%\\
        $\nu_2$ &	319	&16\%		&\bf8\%	&76\%		&100\%\\
        $\nu_3$	&	1462&\bf18\%	&5\%	&77\%		&100\%\\
        $\nu_4$ & 166869&15\%		&5\%	&80\%		&100\%\\\bottomrule
        all		& 164955&14\%		&7\%	&79\%		&100\%
    \end{tabular}
    \caption{Proportion of access modes for each \tb{artist popularity} bin (preferred bins for each access mode are marked in bold).}\label{tab:binmu_types}
    \Description{Table 2: fully described in the text}
\end{table}

\medskip

\begin{figure}[t]
  \centering
  \includegraphics[width=.415\columnwidth]{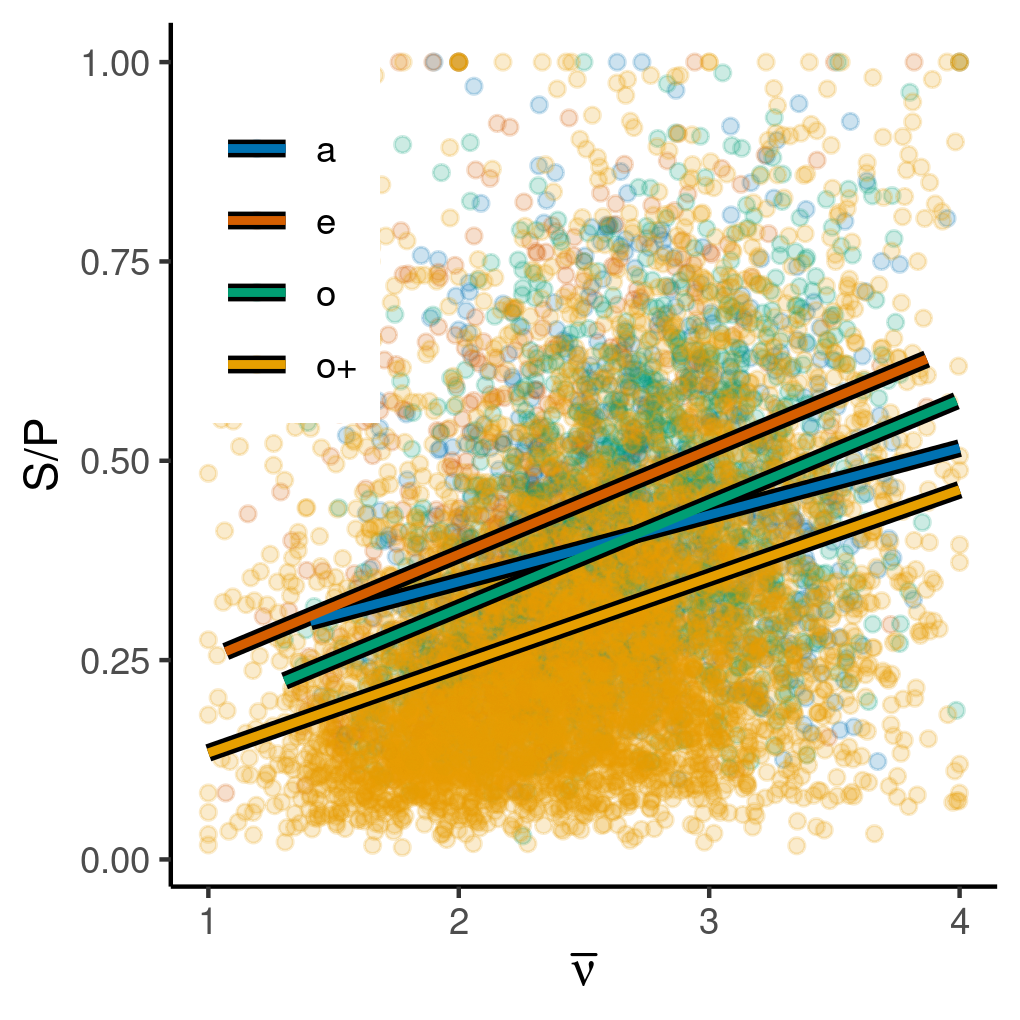}
  \includegraphics[width=.415\columnwidth]{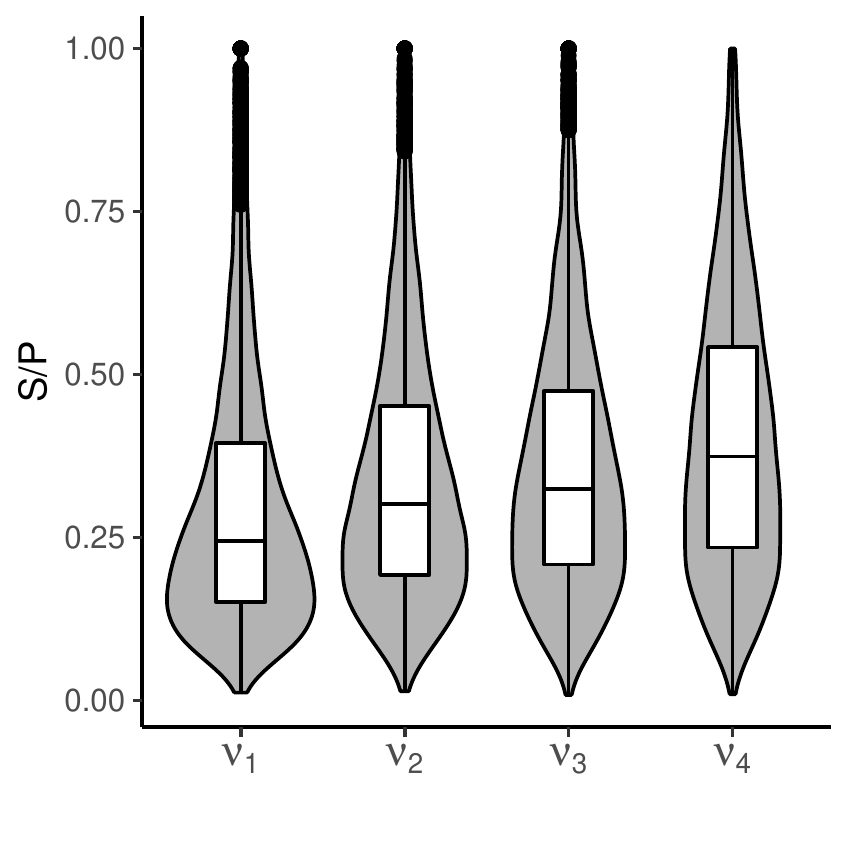}
  \caption{\tb{Artist popularity} and dispersion. {\em Left:} dispersion of listening histories as a function of the average \tb{artist popularity} profile of each user $\bar{\nu}$. Each dot is a user, the solid line represents the best linear fit as a guide to the underlying trend. {\em Right:} boxplot of dispersions restricted to content of a certain \tb{popularity} bin.}
  \label{fig:sp_binmu_sep}  \label{fig:sp_deltamu}
  \Description[Scatterplot and distributions of dispersion versus popularity bin]{Left: users from the four classes are scattered in the dispersion versus popularity plane but an inverse correlation between dispersion and popularity of consumed content is visible for all classes. This can also be seen on the distributions of dispersion for each popularity bin, on the right}
\end{figure}

Finally, we observe that dispersion and \tb{artist popularity} are associated. Part of it is likely mechanical: since the most \tb{popular} bin features much less artists as well as less songs (18,425 songs for $\nu_1$ \hbox{vs.} 1,161,257 songs for $\nu_4$), it also induces a \tb{lower} likelihood of \tb{dispersion}, all other things being equal. Notwithstanding, we essentially observe that users who consume \tb{content by less popular artists} exhibit \tb{more dispersed} listening histories; see \tb{Figure}~\ref{fig:sp_deltamu}-left. 
We refine this picture by computing the dispersion of the listening history of each user restricted to songs of a given \tb{popularity} bin, shown in the right panel of Figure~\ref{fig:sp_binmu_sep}, which also decreases for \tb{content made by popular artists}. 

To summarize, average dispersion increases with activity and decreases with \tb{artist popularity}. Since very organic users have a lower dispersion, this further hints at a positive relationship between the use of recommendation and dispersion.

\subsection{User types and access mode biases}

\begin{figure*}[t]
  \centering
  \includegraphics[width=.96\linewidth]{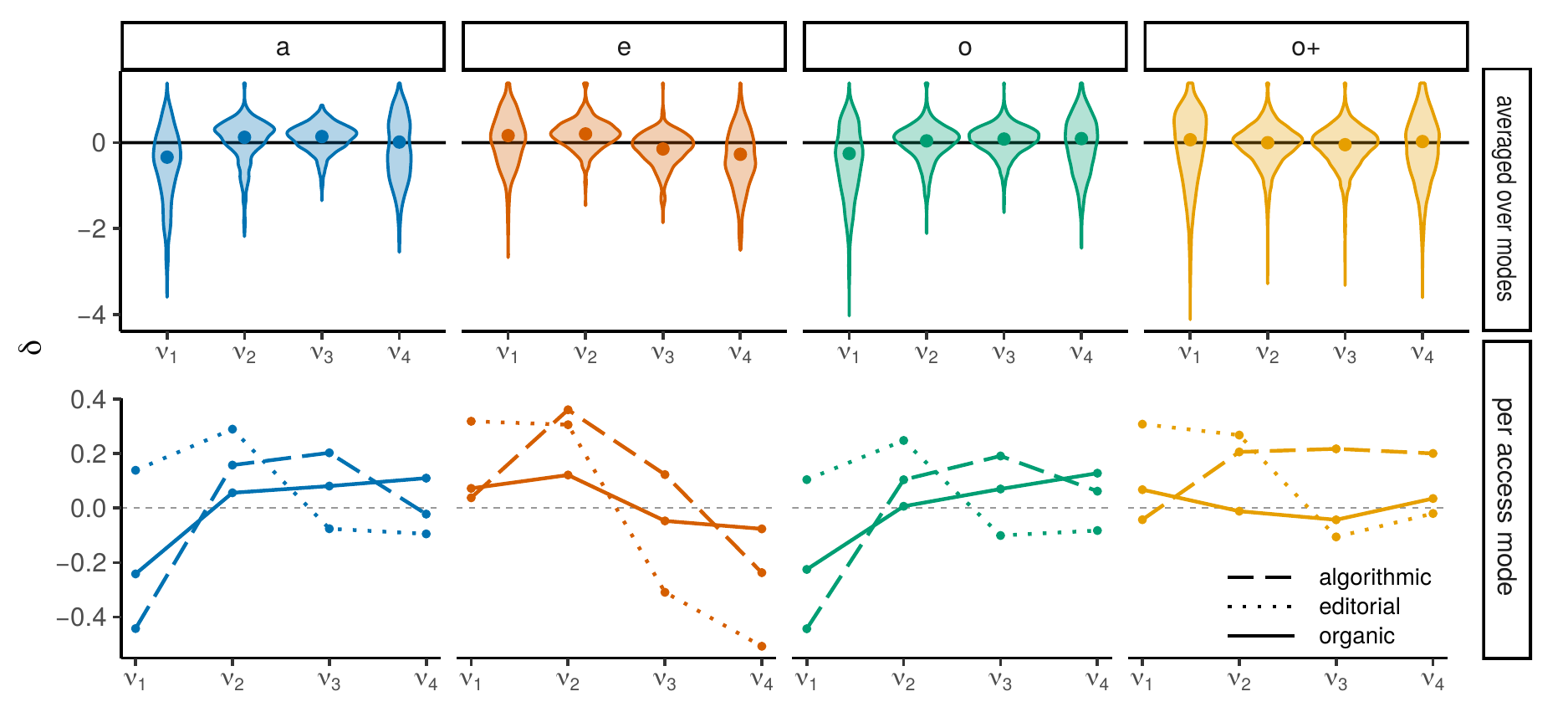}
  \caption{Relative consumption of content from each \tb{popularity} bin, average log-ratio with respect to a uniformly random baseline for each bin (0 corresponds to no deviation, the x-axis is ordered from $\nu_1$ to $\nu_4$ i.e., \tb{for musical content from more to less popular artists}). {\em Top:} average over all plays. {\em Bottom:} breakdown by access \tb{mode}. 
  }\label{fig:binuserlogdelta}\label{fig:binuserlogdeltasep}
  \Description{Figure 5: fully described in the text}
\end{figure*}

We may now appraise the interplay between user types, access modes and diversity from the user perspective. Let us start with \tb{artist popularity}. For each user, we compute the proportion of plays that fall in each \tb{popularity} bin and divide it by its expected value \hbox{i.e.,} we use a null hypothesis where all plays would uniformly fall in equal amounts into each bin (typically proportional to $P/4$ by construction of \tb{popularity} bins). These ratios yield a divergence profile for each user in terms of their \tb{appetency} for some bins \tb{over} others. We then average these profiles over all users of each class and plot in \tb{Figure}~\ref{fig:binuserlogdelta} (top panel) the mean over- or under-consumption per \tb{popularity} bin.

We find that algorithmic users ({\bf a}) \tb{rather} under-consume \tb{content by popular} artists while editorial users ({\bf e}) exhibit an almost opposite profile by over-consuming \tb{such content}. 
Interestingly, {\bf o} and {\bf o+} users markedly differ in their \tb{appetency} for \tb{popular artist} content: whereas very organic users tend to slightly over-listen to \tb{songs from both the most and the least popular artists} \cite[again, consistent with consumer behavior on retail platforms][]{goel-2010-anatomy}, organic users exhibit a monotonous profile favoring less \tb{popular} content in a strictly increasing manner; in this regard, they resemble more algorithmic users. 

We refine this analysis by further plotting over- and under-consumption with respect to access modes. More precisely, given a user type, we perform the normalization locally for each access mode: the ratio between actual and expected quantities is computed using respectively $p_aP$, $p_eP$ and $p_oP$ instead of $P$. The results are shown on the lower panel of figure~\ref{fig:binuserlogdeltasep}. For instance, the curve that corresponds to, say, editorial access of {\bf o+} users, indicates to what extent their editorial plays fall into each \tb{popularity} bins.

We generally confirm that algorithmic access tends to correspond to \tb{content from less popular artists}, even for very organic users ({\bf o+}) who normally over-consume content \tb{from popular artists}, and also for editorial users ({\bf e}) who \tb{normally} consume \tb{even more content from popular artists}, yet to a lesser extent. 
By contrast, on the whole, and for all user types, editorial access pulls towards \tb{popular artists}. Organic access, finally, tends to somewhat mimic average behavior (and expectedly all the more so for the most organic users), but in a less marked manner, suggesting that a good part of the deviation in trends observed on average values stems from recommendation-based access modes, be it editorial or algorithmic.

\medskip
More broadly, these observations contribute to shed light on a typical chicken-and-egg problem \hbox{i.e.,} whether access modes (and thus platform features and affordances) influence user behavior, or user behavior influences how access modes are being used, and for what. In effect, if we focus on organic access as a reference of what users autonomously look for on the platform, we see that user classes exhibit markedly different \tb{appetencies}, for instance with respect to \tb{content from least popular artists}. We also know that user classes roughly entail specialized ways of using the platform, whereby dispersion is lower for the main corresponding access mode (as per Figure~\ref{fig:pdist}), which indicates that the major access mode leans more toward exploitation than exploration. Further, the use of recommendation, algorithmic or not, appears to tilt consumption towards more exploration, all other things being equal. This is further emphasized by the fact that editorial users exhibit a higher dispersion on average than very organic users, even if they generally consume more \tb{popular} content which is otherwise related to lower dispersion in aggregate.  
If we now shift to the use of recommendation-based access modes, we see that they tend to exhibit a general pattern, irrespective of user types: editorial favors \tb{popular}, algorithmic leans toward \tb{least popular}, in very broad and rough terms. In this regard, Figure~\ref{fig:binuserlogdeltasep} contributes to disentangle how both types of effects yield a general consumption profile --- i.e. \emph{both} due to user types and access modes.
It is legitimate to suggest here that we observe and, at least in part, deconstruct the intertwinement of both (1) underlying user types, corresponding to different ways of using the platform, and (2) overlaying access modes, corresponding to the different ways in which the platform may affect user consumption.

\section{The diversity of human-assisted guidance}
Notwithstanding the preferences and expectations of users on the platform, the picture that emerges is that of a split between two types of recommendation in terms of fostering diversity: irrespective of dispersion trends, editors rather than algorithms appear to sustain the consumption of \tb{songs from popular artists} (and vice versa \tb{for least popular artists}). To put this observation in perspective, we need to rely on an external reference. We contend that radio programs constitute a relevant instance in this regard.
Radio-based playlists can be construed as one of the closest offline equivalent of editorial playlists on Deezer --- and, from a human computing viewpoint, one of the oldest large-scale music recommendation systems. 

In practice, it would be difficult to access detailed radio listening histories for a number of people, let alone for the subset of users we considered here. To circumvent this issue, we adjust the way we carry out computations on users to make both sources as comparable as possible. On the one hand, we use artist \tb{popularity} values from the Deezer data set as a general proxy. Around 83\% of songs played on radios are matched with the user data set. These songs inherit their respective artists' \tb{popularity computed} from the Deezer data, and we ignore unmatched songs. By construction, this likely induces an overestimation of \tb{popular} content, and, in turn, of the \tb{popularity bin} for radios that play less \tb{popular} artists.

\begin{figure*}
  \centering
  \includegraphics[width=.46\linewidth]{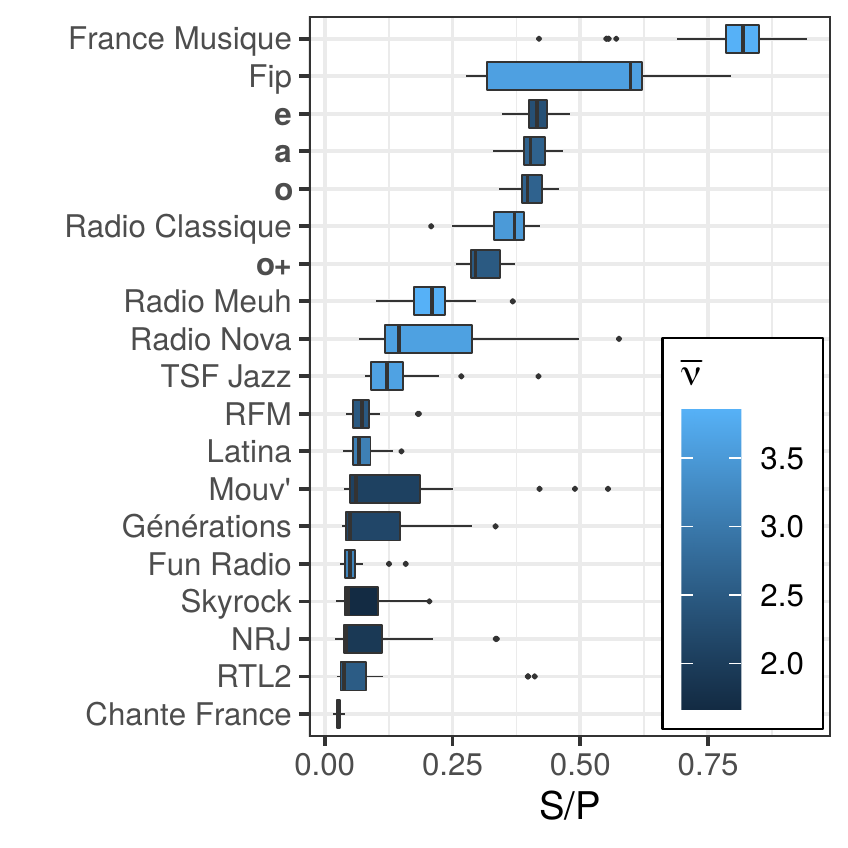}
  \includegraphics[width=.46\linewidth]{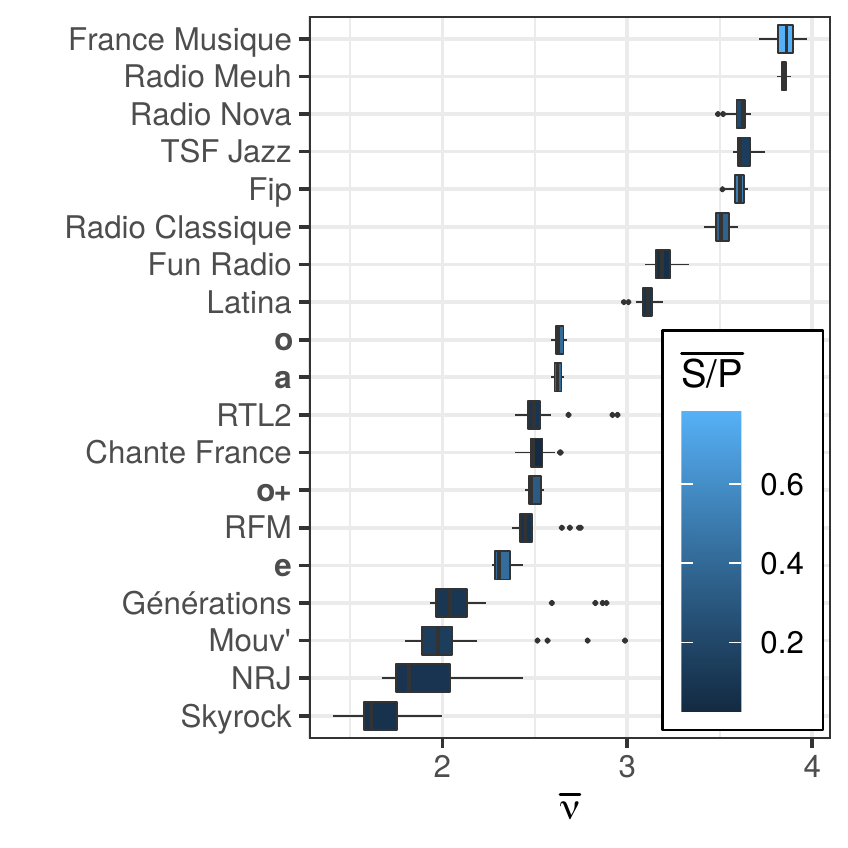}
  \caption{Dispersion and \tb{artist popularity} for a selection of radios and the four user classes. Boxplots are computed on hourly values. {\em Left:} dispersion ranked by decreasing average values, colored by average \tb{popularity} bins. {\em Right:} dually, \tb{popularity} ranked by decreasing average bin number, colored by average dispersion.}
  \label{fig:radios_mu_rates}
  \Description[Diversity measures of average user profiles and radios programs]{Following the method described in full in the text, we see how user's consumed content compares with radio broadcast programs in terms of dispersion and popularity bins. On average, streaming user profiles are on par with the most diverse radios in terms of dispersion (FIP and France Musique) apart from ``o+'' users which are at lower levels. On the artist popularity part, streamers are comparable to more popular stations (including RTL2, Chante France, RFM).}
\end{figure*}

On the other hand, we define hourly listening sessions for both users and radios. This additional focus on hours aims at taking into account the fact that radio playlist may be heavily shaped by the existence of programs broadcast at specific moments of the day, or night. Concretely, for each radio and each user we consider their time-ordered sequence of plays over the year, and we setup counters to keep track of the number of plays $P(h)$ that occurred between hour $h$ and hour $h+1$ during the entire period, along with the number of new songs $S(h)$ that were played during this hour, that is, songs that had never been played previously by this radio (resp. by this user). This way for each radio and user we calculate twenty-four hourly $P$, $S$, $S/P$ and $\bar{\nu}$ values. We then averaged these hourly values over users that belong to the same class, and compare the four user classes previously identified with radio stations, according to the range of their hourly dispersion $S/P$ and hourly average \tb{popularity} $\bar{\nu}$ values.

Results are gathered on figure~\ref{fig:radios_mu_rates}. We select a representative sample of 15 radios out of 39 for clarity purposes. On the left panel, we compute dispersion boxplots both for radios and user types. All items, radios or user types, are ordered from top to bottom by decreasing value of average dispersion. We further color boxplots according to the average \tb{popularity} bin. Table~\ref{tab:examples} additionally gathers the detailed breakdown of plays in each \tb{popularity} bin, for all user types and a few selected broadcasters.

We observe that most user types exhibit more dispersion than most radios, indicating that radio programming is  tilted toward the exploitation of a more limited catalog. More precisely, there appears to be an inflection point around ``TSF Jazz'' and ``RFM'' which roughly splits the set of items between a larger half with a significantly low dispersion (below 0.10) and a smaller half with much higher dispersions (generally above 0.25). In this picture, user classes exhibit among the highest values.  Remarkably, two radios are above all user classes --- France Musique and FIP, which are public-funded, predominantly musical and rather eclectic broadcasters.   Besides, a mild correlation between \tb{popularity} and dispersion is visible: the diversity of the catalog in terms of distinct played items is roughly linked to its diversity in terms of playing \tb{artists from the less popular bins}.

\begin{table}\centering\small
	\begin{tabular}{r|rrrr|c}\toprule
	&\multicolumn{1}{c}{$\nu_1$}&\multicolumn{1}{c}{$\nu_2$}&\multicolumn{1}{c}{$\nu_3$}&\multicolumn{1}{c|}{$\nu_4$}&$\langle S/P\rangle$\\
        \midrule
        \bf a&  17\%&  27\%& \bf 29\%&  27\%&0.39\\
        \bf e&  29\%&  \bf 30\%&  22\%&  19\%&0.41\\
        \bf o&  22\%&  25\%&  26\%&  \bf 27\%&0.38\\
        \bf o+&  \bf 29\%&  23\%&  23\%&  25\%&0.28\\
        \midrule
        Radio Meuh  &   0\% &  2\% & 16\% &\bf 82\%& 0.23 \\
        FIP			&	2\%	&	7\%	&20\%	&\bf 71\%	&0.60\\
        RFM			&  7\%&  27\%&\bf 36\%&  30\%&0.10\\
        Fun Radio& 18\%& \bf 61\%& 13 \%&8\%&0.04\\
        NRJ			&  \bf41\%&  23\%&  23\%& 7\%&0.03\\
		\bottomrule
    \end{tabular}
    \caption{Average breakdown of content played into each \tb{popularity} bin, for all user types and a selection of radios, and average dispersion.
    }\label{tab:examples}
    \Description[Dispersion and popularity vary a lot depending on user types and radios]{Radios differ a lot in terms of diversity. FIP has a dispersion of 0.6 while Fun Radio has just 0.04. Radio Meuh plays the least popular content with more than 80 per cent coming from the last bin while NRJ has more than 41 per cent from the first one. Prototype users for the four classes are in-between these extremes.}
\end{table}

A dual computation consists in representing boxplots of observed mainstreamness averaged over hours, while coloring them with the average dispersion --- see the right panel of \tb{Figure}~\ref{fig:radios_mu_rates}. While the correlation between dispersion and popularity is not immediately visible, the Spearman rank correlation between S/P and the proportion of songs in $\nu_4$ is equal to 0.827, confirming the similar \tb{inverse} ordering of dispersion and popularity: \tb{radios with more dispersed programs also play more songs from less popular artists}. There are notable exceptions such as ``Radio Meuh'', whose lower bound on popularity is the highest of all radios ---and it is indeed considered informally as a very eclectic broadcaster--- while having a relatively moderate dispersion (as seen in Table~\ref{tab:examples}). The new interesting take-away of this second plot lies in the positions of user types. They are now much closer to the median value of this selection, even though organic and algorithmic users appear to be a little bit above it.  Put differently, while online music listening practices seem to foster functional diversity (dispersion), in terms of semantic diversity (\tb{popularity}) they seem, on average, to be neither significantly above nor significantly below the mass of the radios we focus on.

\section{Concluding remarks}

Our analysis of recommendation, in the broad sense, on a music streaming platform started \tb{by} assuming the existence of distinct attitudes towards platform affordances. 
This produced a typology of four user classes who \tb{markedly} differ in their modes of access to music and, more importantly, in the way these modes are associated with distinct effects on the diversity of consumed songs. We showed more broadly that there is no blanket answer to the question of the \tb{influence} of recommendation \tb{devices: rather, we contend that it primarily} depends on users. This observation may have important ramifications in the appraisal of the effect of recommender systems on consumption diversity --- we suggest to speak of filter niches, rather than bubbles.

Moreover, we showed that the framing of recommendation on platforms and, thus, their algorithmic governance, would benefit from paying specific attention to the various types of guidance. Human-mediated curation is associated with distinct effects on the exploration of content and fostering diversity, and with distinct user classes. Both types of guidance seem to often have opposite effects. They may also fulfill distinct roles and thus have to obey to distinct design principles.  We further introduced a bridge between platform-based recommendation and traditional offline curation, exemplified by radio playlists, which we see as a fruitful point of comparison to frame how platforms may or may not affect user preferences and access to content. If user classes were radios (and assuming it is valid to speak of an average user for each user class), they would certainly not appear to be more repetitive than most radio stations, while they would exhibit a wide range of serendipity. Interestingly, while algorithmic access modes seem to generally be associated with an avoidance of \tb{popular} content, it is what we call organic users who appear to focus most on the least \tb{popular} content --- yet, these users precisely exhibit a relatively balanced diet of platform affordances, generally combining their autonomous navigation with both sorts of recommendation. In other words, we may hypothesize that the most ``expert'' users best exploit the platform capabilities to explore the long tail. Editorial access, on the other hand, and even more so for editorial users, appears to fulfill a role traditionally ascribed to mainstream radios --- putting forward a higher proportion of \tb{popular} artists.

A step further from this study would consist in analyzing temporal aspects of the partitioning. Indeed, the chronology of access modes to a song probably carries valuable information. For instance, a song may first be discovered through an editorial recommendation, then later accessed organically and symmetrically, organically accessed content may be subsequently picked by the algorithm later on. More generally, user trajectories and transitions between classes should be the focus of future research that would shed light on the acclimation of users to platform features. In particular, the causal relationship that may exist between the presence or absence of some content in editorial and algorithmic recommendations and the evolution of their platform-wide \tb{popularity} would also be a fruitful field of investigation.
With respect to the ubiquitous debate on whether platform recommendation fosters diversity or not, a mixed picture emerges depending on which types of recommendation one is talking about, as well as which type of audience --- whereby radios provide an insightful reference.
On the whole, this comparison may shed light on the relative position of platforms and their contribution to cultural diversity with respect to traditional cultural prescribers and, again, how their algorithmic principles should be governed. Understanding user expectations not only in terms of genres and tastes, but in terms of their use and perhaps understanding of the technical features of a platform, may help refine the design of recommender systems ex ante, rather than optimizing their results ex post. 
\tb{The corresponding takeaway for practitioners is to be attentive to the differences between access modes which may fulfill distinct roles for listeners \hbox{e.g.,} active exploration \hbox{vs.} background music listening.}
 Longitudinal analyses of the evolution of user tastes, specifically in terms of learning and of long-term effects of distinct platform affordances, as well as qual-quant analyses based on user surveys and interviews would be most helpful in further disentangling the chicken from the egg.

\begin{acks}
This paper has been partially realized in the framework of the “RECORDS” grant (ANR-2019-CE38-0013) funded by the ANR (French National Agency of Research). We are grateful to Dougal Shakespeare for useful comments.
\end{acks}



\end{document}